# High frequency nano-optomechanical disk resonators in liquids

E. Gil-Santos[1]*, C. Baker[1], D. T. Nguyen[1], W. Hease[1], A. Lemaître[2], S. Ducci[1], G. Leo[1] and I. Favero[1]*

[1] Matériaux et Phénomènes Quantiques, Université Paris Diderot, CNRS, Sorbonne Paris Cité, UMR 7162, 10 rue Alice Domon et Léonie Duquet, 75013 Paris, France.

[2] Laboratoire de Photonique et Nanostructures, CNRS, Route de Nozay, 91460 Marcoussis, France.

*eduardo.gil-santos@univ-paris-diderot.fr

*ivan.favero@univ-paris-diderot.fr

**Abstract**

Vibrating nano- and micromechanical resonators have been the subject of research aiming at ultrasensitive mass sensors for mass spectrometry, chemical analysis and biomedical diagnosis. Unfortunately, their merits diminish dramatically in liquids due to dissipative mechanisms like viscosity and acoustic losses. A push towards faster and lighter miniaturized nanodevices would enable improved performances, provided dissipation was controlled and novel techniques were available to efficiently drive and read-out their minute displacement. Here we report on a nano-optomechanical approach to this problem using miniature semiconductor disks. These devices combine mechanical motion at high frequency above the GHz, ultra-low mass of a few picograms, and moderate dissipation in liquids. We show that high-sensitivity optical measurements allow to direct resolve their thermally driven Brownian vibrations, even in the most dissipative liquids. Thanks to this novel technique, we experimentally, numerically and analytically investigate the interaction of these resonators with arbitrary liquids. Nano-optomechanical disks emerge as probes of rheological information of unprecedented sensitivity and speed, opening applications in sensing and fundamental science.

The development of ultrasensitive mass sensors for biological applications like early disease detection has recently generated a great deal of effort[1-3]. In this context, micro and nanomechanical resonators with low inertial mass appear as a key technology, as exemplified by their capability to sense down to individual atoms in vacuum[4-6]. In short, the minimum detectable mass is proportional to the effective mass of the resonator and sensitivity improves if mechanical dissipation is reduced. Device miniaturization and dissipation control are therefore crucial. In liquid -typical of biological environments-, mechanical energy losses are high and the mass sensitivity diminishes dramatically[7,8]. Furthermore, viscous damping in a liquid also increases when standard mechanical devices such as cantilevers or membranes are miniaturized[9].

To circumvent these problems novel structures and techniques have been developed. An efficient approach is to use vibrating microchannel cantilevers where the liquid flows directly inside the resonator in order to reduce the induced dissipation [10-12]. Fibered microcapillaries[13] adopt this approach too with a more integrated optical detection. Since both types of devices need to embed fluidic channels, their size varies from a few tens to few hundreds of microns, leading to masses above the nanogram and mechanical frequencies at most in the MHz range. These channel devices can hardly be miniaturized to the nanoscale. Other geometries have been investigated, including the case of a partially immersed resonator[14], which faces similar size limitations and has demonstrated lower integration capabilities.

Here we put into light the potential of a nano-optomechanical approach in this context, focusing in particular on miniature (sub-µm$^3$ volume) semiconductor disk resonators and their (contour) mechanical Radial Breathing Modes (RBMs). These resonators can be completely immersed in a liquid, allowing in situ fluidic operation; and they have dimensions compatible with multiple detection sites on a single chip. While their low inertial mass (picogram), high mechanical frequencies beyond the GHz and moderate fluidic dissipation are clear assets for sensing applications; they have not been investigated in liquids so far[15,16]. Indeed the typical mechanical displacement associated to RBMs is extremely small and difficult to detect with conventional techniques, especially in a liquid and at very high frequencies involved. Here we show that nano-optomechanics[17,18] provides the required bandwidth and precision to face this experimental challenge. Indeed in ambient conditions miniature disks support optical Whispery Gallery Modes (WGMs) of high optical quality factor ($Q>10^5$) that strongly couple to RBMs[19], yielding a displacement sensitivity of $10^{-17}$ m/√Hz at GHz frequencies[15]. Exporting these features, we achieve resolving optomechanically the Brownian motion of disk resonators at very high frequency (GHz), even in the most dissipative liquids. The technique allows studying with great detail the fluid-structure interaction. Experimental results are interpreted in the light of novel models that directly grasp the role of physical parameters involved in viscous and acoustic interactions. They lead us to quantify the unprecedented sensitivity of nano-optomechanical disk resonators to mass deposition and other rheological changes in liquids, setting the stage to a variety of sensing and basic science applications at very high frequency.

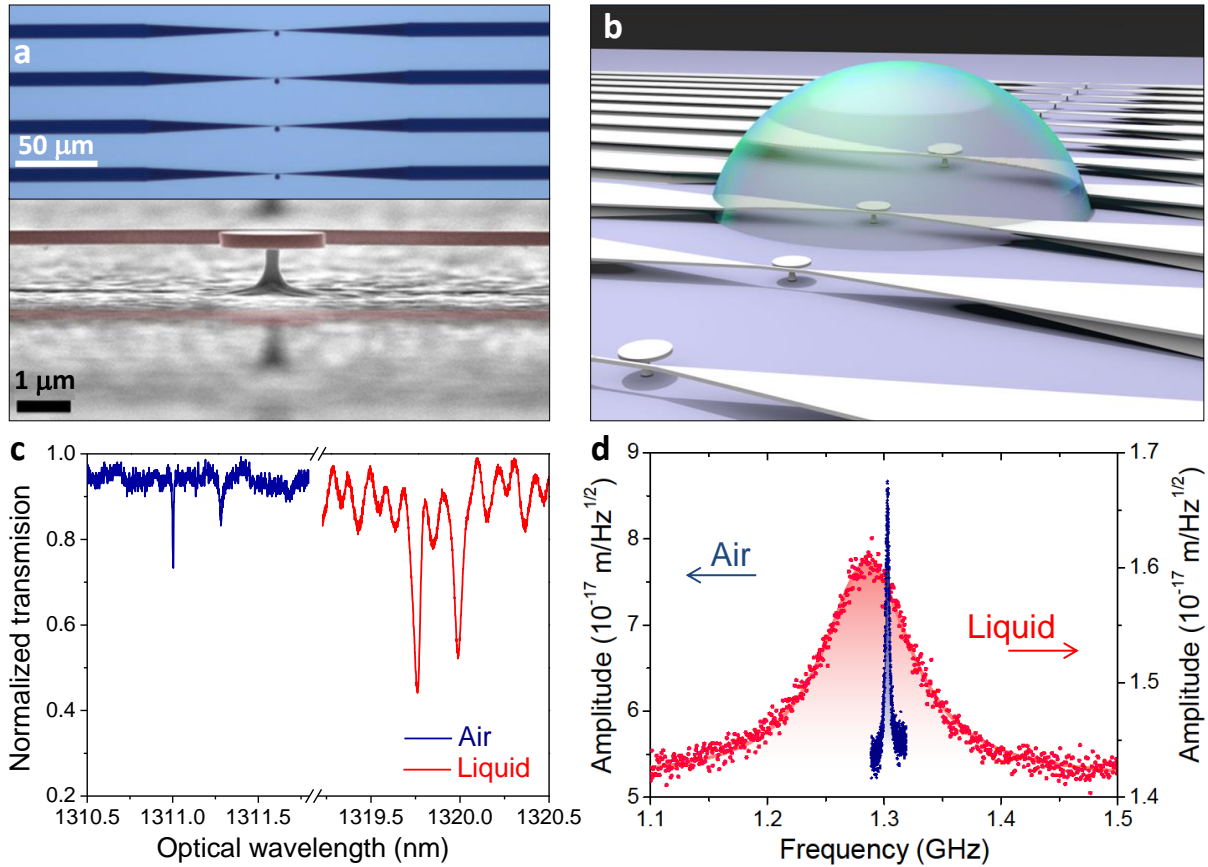

**Figure 1: Operation of nano-optomechanical disks in liquid. a,** Upper panel: Optical top view of 4 waveguide-disk structures with taper. Lower Panel: scanning electron microscope side-view of a disk resonator sitting on its pedestal, close to the suspended tapered part of the waveguide. **b,** Schematic representation of resonators immersed in a liquid droplet. **c,** Optical transmission spectrum of a 1μm radius disk in air (dark blue) and in the liquid (red). **d,** Thermomechanical spectrum of the disk vibrating in air (dark blue) and in the liquid (red), acquired by optomechanical measurement.

In this work, miniature disk resonators and their pedestal are fabricated out of Gallium Arsenide (GaAs) and Aluminum Gallium Arsenide (AlGaAs), whose epitaxial control provides low-dissipation high purity crystal and the possibility of engineering optomechanical heterostructures[20,21]. However, the general implications of our work extend as well to other mature semiconductor platforms with similar refractive properties, such as silicon. In our experiments, the disk thickness is kept constant (320 nm) while the disk radius varies from 1 to 3 µm. Monochromatic light at $\lambda=1.3$ µm is evanescently coupled into the disk through a GaAs tapered waveguide fabricated on chip in the disk's vicinity[22]. A chip is typically millimeter-sized and contains several waveguides/disk units (see Fig. 1a). Thanks to the large refractive index of GaAs, the disk/waveguide structures can be immersed in a transparent liquid while still preserving their light guiding properties, i.e. leaving their overall transmittance unaffected. In what follows, a microliter droplet of liquid, deposited on the sample surface with a micropipette (see Fig. 1b), covers many resonators and lasts from few to hundreds of minutes before evaporating. After immersion, the disk WGM resonances are red-shifted and slightly broadened, as apparent in Fig. 1c. These effects result from the refractive action of the surrounding liquid, as confirmed by FEM simulations (not shown). For a radius of 1 µm, Fig. 1c reports a loaded optical Q of $2.3\times10^4$ in the liquid, which combined with a strong optomechanical coupling $g_0=2$ MHz[19] leads to a high motional sensitivity down to $10^{-17}$ m/√Hz. Such sensitivity is sufficient to resolve the minute thermal (Brownian) fluctuations of the mechanical system, even in the liquid. Fig. 1d shows mechanical spectra obtained by such optomechanical measurements in the Brownian motion regime before and after immersion. Here again, the effect of immersion is twofold: a red shift of the mechanical frequency resembling an added motional inertia, and a broadening of the resonance corresponding to added dissipation. We now provide an in-depth analysis of the dispersive and dissipative facets of the fluid-structure interaction.

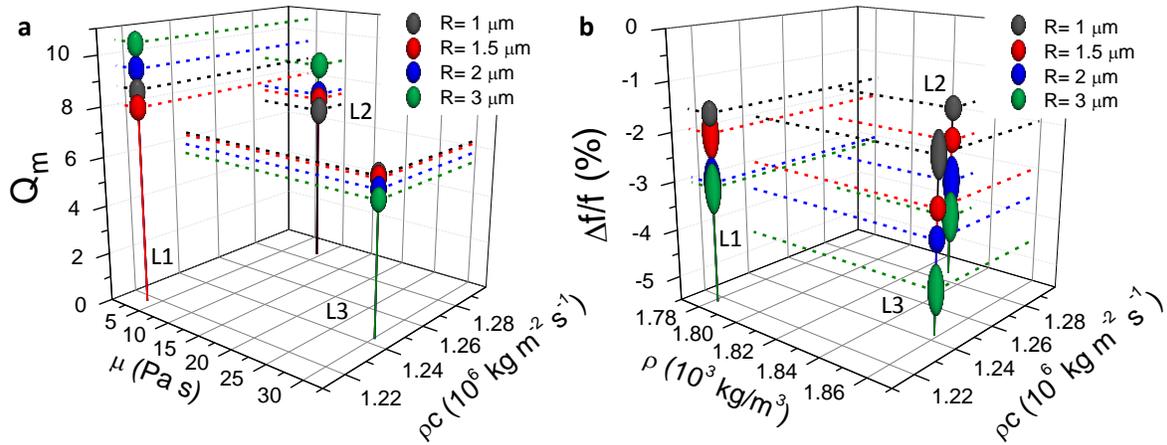

**Figure 2: Dissipative and dispersive fluid-structure interactions measured by nano-optomechanical means. a,** Mechanical quality factor $Q_m$ measured for four distinct disk radii in three different liquids, plotted versus the viscosity and the acoustic impedance of the liquid. **b,** Relative mechanical shift $\Delta f/f$ for the same set of measurements, plotted versus the density and the acoustic impedance of the liquid. Symbol vertical size corresponds to error bar.

The mechanical behavior of a GaAs disk in a fluid is expected to depend both on the disk dimensions and fluid properties, motivating the measurement of resonators with different radius immersed in different liquids. Fig. 2 reports the measured mechanical $Q_m$ and relative mechanical frequency shift $\Delta f/f$ of disks with radius varying from 1 to 3 µm, immersed in three remarkably distinct liquids whose properties such as dynamic viscosity µ, density ρ and speed of sound c are listed in Table 1. These measurements are carried out in the low optical power limit as discussed in the supplementary notes. While our goal below is to interpret these results, we remind first that the general problem of a mechanical element interacting with an arbitrary liquid is extremely complex, formally merging elasticity theory and generalized Navier-Stokes equations. Exact solutions are certainly out of reach, but some considerations are possible in simplified hydrodynamic situations. For an incompressible viscous liquid, the disk motion dissipates energy through fluidic friction, in proportion to the viscosity parameter µ (viscous regime). Added motional inertia occurs as well in link to the displaced mass of liquid, in proportion to ρ. In the non-dissipative compressible liquid case (acoustic regime), the moving disk generates outgoing pressure waves in the liquid leading to dissipation and modified vibration frequency. To grasp these different aspects of the fluid-structure interaction, Fig. 2 plots $Q_m$ as a function of viscosity and acoustic impedance ($\rho c$) of the liquid, and $\Delta f/f$ versus density and acoustic impedance. Under such representation, some trends appear in the measurements. Firstly, whatever the size of the disk, $Q_m$ decreases as $\mu$ increases, as expected from viscous interactions. Secondly, the negative magnitude of $\Delta f/f$ increases as $\rho$ increases, like anticipated from an added mass picture. The behavior as a function of the acoustic impedance is less obvious. $Q_m$ and $\Delta f/f$ are non-monotonous functions of $\rho c$, suggesting impedance matching conditions that will be elucidated below. Looking at the magnitude of these effects, our measurements suggest that both viscous and acoustic interactions must be taken into account in order to accurately describe the behavior of nano-optomechanical when operation in liquid. This is in contrast to lower frequency mechanical systems, which tend to dwell in the viscous regime for common fluids like water or air [9,23]. Below, consistent numerical and novel analytical models of viscous and acoustic interactions are proposed and utilized to interpret experimental results. Analytic expressions lead us to reveal the potential of nano-optomechanical disks as sensors of mass deposition and other rheological changes in liquids.

|  | $\rho$ (kg/m$^3$) | $\mu$ (mPa s) | c (m/s) |
|---|---|---|---|
| Water | 1000 | 1 | 1500 |
| Liquid 1 | 1782 | 3.5 | 669 |
| Liquid 2 | 1831 | 9 | 702 |
| Liquid 3 | 1859 | 30 | 683 |

**Table 1: Properties of liquids.** The density $\rho$, dynamic viscosity $\mu$ and speed of sound c of are quoted at room temperature. Liquids 1, 2 and 3 are perfluorinated liquids of growing viscosity and density.

A key feature of optomechanical disks vibrating on their RBM is that they possess rotational invariance around their vertical axis, provided their constitutive elastic material is considered isotropic. In an isotropic liquid, this symmetry allows to reduce the difficult 3D fluid-structure interaction problem to an easier 2D. Finite Element Method (FEM) then allows performing reliable numerical calculations of the disk vibrating in the liquid, be it in the purely incompressible viscous or purely acoustic regime. Details about these FEM calculations, which provide a first analysis tool for the present work, are given in the supplementary notes.

But the rotational symmetry has other consequences. For a small-radius pedestal, the RBM of the disk can be approximated by the radial contour extension of a circular plate, which admits analytical solutions[19,24,25]. By means of the Poisson effect, this extensional mode is accompanied by compression ("pinching") in the disk's thickness, which can also be described analytically with a high precision (supplementary notes). We build upon this analytical description to develop a theory of the fluid-disk interaction. To that purpose, we pave the vibrating disk surface with elementary vibrating spheres interacting with the fluid, and then sum the contribution associated to all spheres[23,26]. Indeed, due to its perfect symmetry, the sphere vibrating harmonically along one direction in a viscoelastic liquid is a problem that admits an exact solution, which was derived by Oestreicher [27]. In this derivation, the complete velocity field in the fluid was solved, allowing expressing the fluidic force exerted on any point of the sphere surface. In the disk RBM case, we make use of these point-dependent expressions to account precisely for the response of the liquid against the radial and the out of plane "pinching" motion of the disk, finally encompassing dissipative and dispersive fluid-structure interactions with a very satisfactory precision.

Let us first analyze the incompressible viscous regime of the liquid. In that case, Oestreicher's formulas for the viscous force are integrated on our analytical profile of the RBM and lead the following expressions for the mechanical quality factor and the resonance frequency shift of the disk in the liquid (supplementary notes):

$$Q_{viscous} = \frac{\rho_s \omega H R}{8.36 \cdot \mu + (3.18 \cdot H + R)\sqrt{2\rho\omega\mu}} \quad (1)$$

$$\left(\frac{\Delta f}{f}\right)_{viscous} = -\frac{1.27 \cdot H}{2R}\frac{\rho}{\rho_s} - \frac{3.18/R + 1/H}{2\rho_s}\sqrt{2\rho\mu/\omega} \quad (2)$$

where $R$, $H$, $\rho_s$ are the radius, thickness and density of the disk material respectively. In these formulas, we see that dissipation not only stems from the fluid viscosity µ but also from a dynamical term $\sqrt{\rho\omega\mu}$, which is actually dominant in the frequency regime of nano-optomechanical disks. Given the frequency expression for the first-order RBM of a circular plate $\omega = \frac{\lambda}{R}\sqrt{\frac{E}{\rho_s(1-\nu^2)}}$ (with E and ν the Young modulus and Poisson ratio of the material[19]), where the frequency parameter λ only depends on ν, we can obtain the following scaling law from equation (1): $Q_{viscous} \propto \frac{H\sqrt{R}}{3.18H+R}$. Accordingly Q has a maximum at $R = 3.18H$ and increases with the thickness. An important consequence is also that Q only diminishes by a factor 10 when $H$ and $R$ are reduced by a factor 100, bringing a reduction by a factor $10^6$ in the device mass. In this view, miniaturization of disks vibrating on their RBM appears appealing for Q-dependent applications, at variance with standard cantilever devices. On the dispersive side of the fluid-structure interaction, the mechanical frequency shift is proportional to the fluid density but also to a dynamical term $\sqrt{\rho\mu/\omega}$, both having commensurable contributions for our miniature disks.

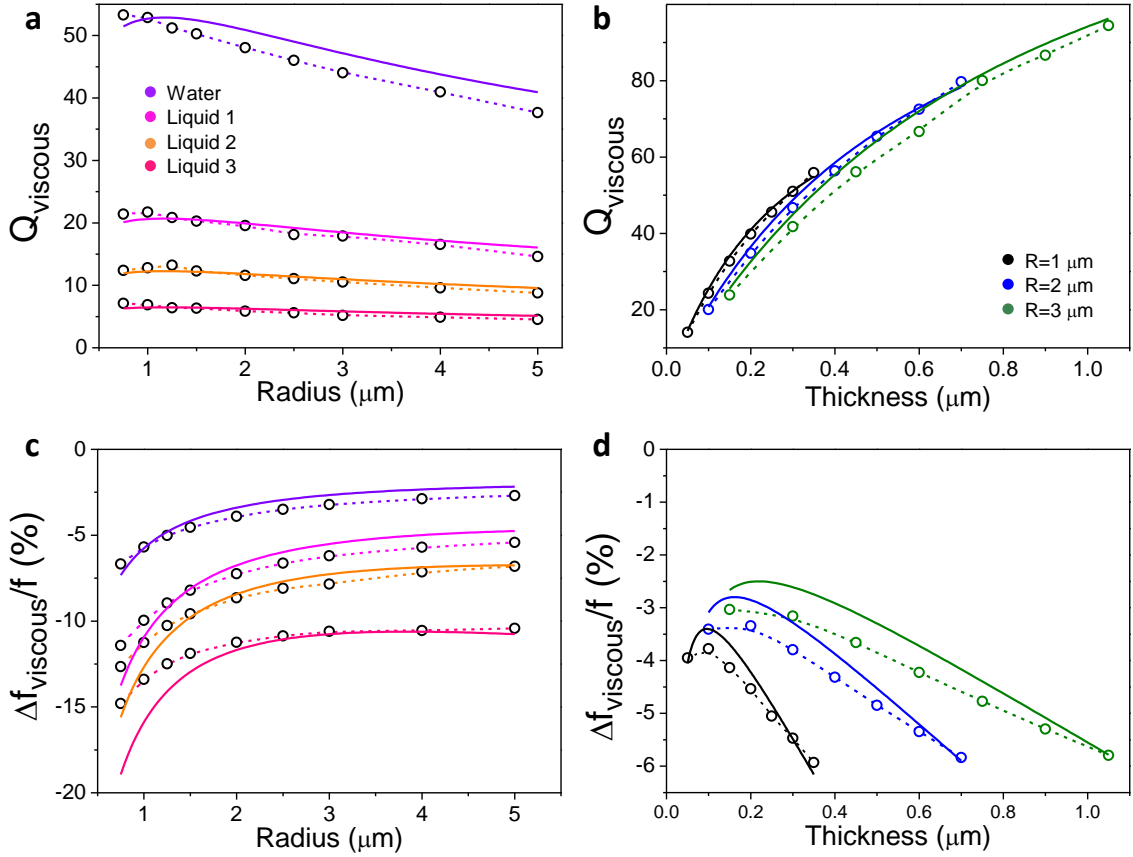

**Figure 3: Viscous regime models.** Our analytical model (solid lines) is systematically compared to FEM results (open circles with dashed lines). **a, b,** Mechanical quality factor versus **(a)** the disk radius for a fixed thickness of 320 nm in four distinct liquids. **(b)** disk thickness for three distinct radius when immersed in water. **c, d,** The corresponding relative mechanical shift Δf/f as a function of same parameters.

Our analytical formulas are compared to numerical FEM simulations in Fig. 3, for the four different liquids considered in this work. For all liquids, both the dissipative and dispersive parts of the fluid-structure interaction are reproduced with a remarkable level of agreement, as a function of the disk radius and thickness. The Q maximum close to $R = 3.18H$ if for example directly visible in FEM. An excellent agreement as a function of liquid's viscosity and density is also illustrated in the supplementary notes.

Let us now analyze the acoustic regime. Following the same path as above, the fluidic force is integrated over the disk moving surface to obtain the mechanical Q and frequency shift of the RBM in the fluid. Although this analytical approach reproduces qualitatively many aspects observed in FEM simulations, in the acoustic regime it fails in a more quantitative fit. This problem stems from interferences of compression waves emitted by the disk in the liquid, whose detailed patterns are not retrieved correctly by the spheres approach. Still the obtained analytical expressions are an excellent starting point to develop more accurate empirical formulas. Guided by these analytical expressions and by FEM calculations, we finally obtain the following empirical formulas (supplementary notes):

$$Q^{-1}_{acoustic} = \frac{H}{R}\frac{\rho}{\rho_s}\left[1.92\frac{\left(\frac{1}{2\sqrt{1-\nu^2}}\frac{\lambda}{c}\frac{c_s}{c}\right)^3}{4+\left(\frac{1}{2\sqrt{1-\nu^2}}\frac{\lambda}{c}\frac{c_s}{c}\right)^4} + 1.05\left(\frac{H}{R}\right)^{-1/2}\frac{\left(\frac{3}{4\sqrt{1-\nu^2}}\frac{\lambda}{c}\frac{c_s H}{c R}\right)^3}{4+\left(\frac{3}{4\sqrt{1-\nu^2}}\frac{\lambda}{c}\frac{c_s H}{c R}\right)^4}\right] = f(\frac{H}{R},\frac{\rho}{\rho_s},\frac{c_s}{c},\nu) \quad (3)$$

$$\left(\frac{\Delta f}{f}\right)_{acoustic} = -0.954\frac{H}{R}\frac{\rho}{\rho_s}\left[\frac{1}{4+\left(\frac{1}{4\sqrt{1-\nu^2}}\frac{\lambda}{c}\frac{c_s}{c}\right)^4} + \frac{2.4}{4+\left(\frac{3}{5\sqrt{1-\nu^2}}\frac{\lambda}{c}\frac{c_s H}{c R}\right)^4}\right] = g(\frac{H}{R},\frac{\rho}{\rho_s},\frac{c_s}{c},\nu) \quad (4)$$

Importantly, the two expressions can be cast into simple functions of dimensionless parameters H/R, $\rho/\rho_s$, $c/c_s$ and $\nu$. Regarding dissipation, the dependence of Q on the geometric factor H/R shows that the disk mass can be miniaturized at constant H/R while maintaining constant Q, leading here again to improved performances with miniaturized devices. A second lesson brought by these formulas is that the density and speed of sound ratios $\rho/\rho_s$ and $c/c_s$ each play a separate role, in contrast to the standard case of acoustic plane wave propagation that only involves the acoustic impedance ratio $\rho c/\rho_s c_s$. As far as mechanical shift is concerned, the linear dependence with $\rho$ supports the picture of an added mass by the liquid. Noteworthy is also the scaling relation $\Delta f/f \propto H\rho/R\rho_s$ that was already present in the viscous case.

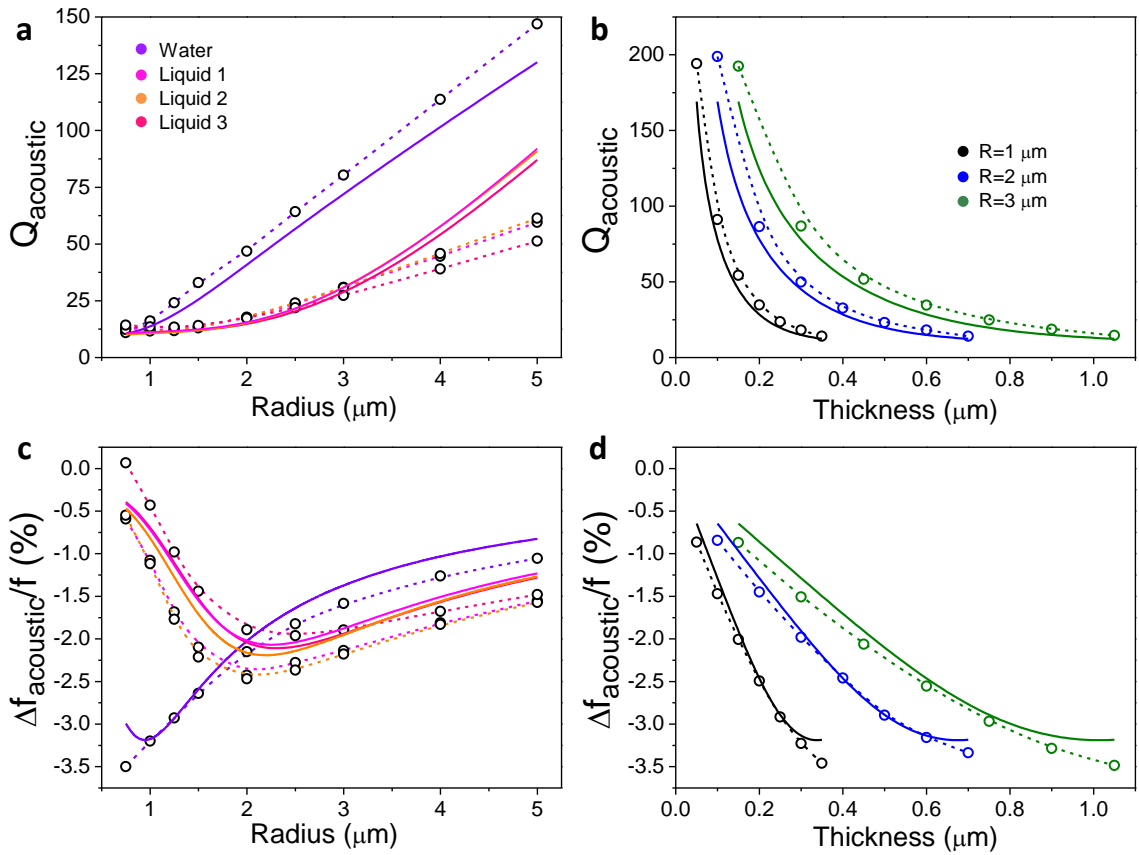

**Figure 4: Acoustic regime models.** Our analytical model (solid lines) is systematically compared to FEM results (open circles with dashed lines). **a, b,** Mechanical quality factor versus (**a**) disk radius for a fixed thickness of 320 nm in the same 4 distinct liquids as above, (**b**) disk thickness for three distinct radius when immersed in water. **c, d,** The corresponding relative mechanical shift Δf/f as a function of same parameters.

These analytical acoustic formulas are compared to numerical FEM simulations in Fig. 4 for the four liquids. For all liquids, the evolution of Q and Δf/f as a function of the disk's radius and thickness is reproduced satisfactorily, with quantitative discrepancies limited to below 10%. A remarkably solid agreement is also demonstrated in the supplementary notes as a function of liquid's density and speed of sound.

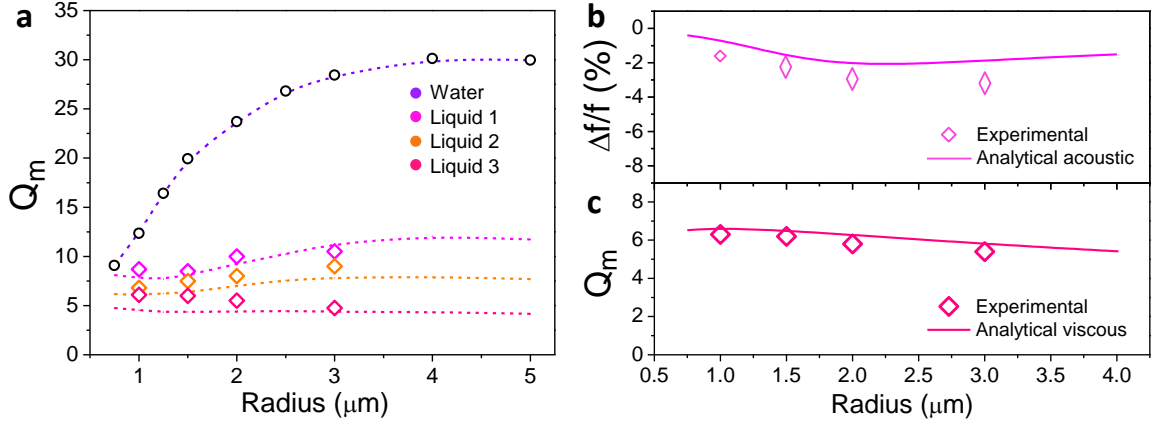

**Figure 5: Interpretation of nano-optomechanical experiments in liquids. a,** Radius-dependent mechanical quality factor measured in liquids (open diamond symbols) and fit obtained by summing viscous and acoustic dissipation calculated by FEM (dashed lines). A solid agreement is observed over the whole range of parameters. The most precise fit is obtained for the least-viscous water-like liquids. FEM results for water are shown in open circles linked by a violet dashed line. **b,** Radius-dependent mechanical shifts measured in liquid 1 fitted by the analytical acoustic model. **c,** Radius-dependent mechanical quality factor measured in liquid 3 and fitted by the analytical viscous model.

With these novel models in hands, let us now interpret our experimental results. Figure 5a reports the mechanical Q of disks with radius varying between 1 and 3 μm, measured in liquids 1, 2 and 3. The measurements are satisfactorily fitted by both our FEM and analytical models, when summing the dissipation associated to viscous and acoustic interactions. The model predicts a Q reaching 30 in water for a radius of 4 μm. At 1μm, the acoustic dissipation dominates and sets Q=12, leading to a remarkable Q-frequency product Qf of $1.8\times10^{10}$ in water. The minimum detectable mass can be evaluated using the well-established formula[28,29] $\delta m_{min} = 2m\sqrt{\frac{B}{2\pi Qf}\frac{k_BT}{E_{max}}}$ where m is the effective motional mass of the device, B the measurement bandwidth and $E_{max}$ the maximal mechanical energy stored in the resonator. For contour-modes, $E_{max}$ is limited by the intrinsic mechanical non-linearity of the material[30,31]. In silicon, the non-linearity settles at an energy density $E_{max}/V$ of $2\times10^5$ J/m$^3$ for a bulk acoustic wave resonator with $Q>10^5$, extrapolating to $2\times10^9$ J/m$^3$ for the present Q=12, a value that should not differ in GaAs[32]. With these parameters, the minimum mass detectable for an integration time of 1s is 14 yoctogram. Mass detection at this level was only demonstrated in experiments employing carbon nanotubes at cryogenic temperature under vacuum[6]. While recent vibrating-channel devices in water approached the 10 zeptogram range [33], the yoctogram regime remains unexplored in liquids. To further explore the potential of nano-optomechanical disks as sensors of fluidic information, we focus in Fig. 5 b on the mechanical shifts measured in liquid 1. This is the least viscous of the three perfluorinated liquids, such that the acoustic model alone suffices to explain experimental results. Equation (6) is hence employed to evaluate the sensitivity of the device to density changes in moderately viscous liquids like water. For the same disk as above placed in water, it leads to an approximate scaling with density $\Delta f \propto \rho$, with little role played by changes in speed of sound. The thermodynamical limit of detection of density changes amounts to $\delta\rho_{min}=2\times10^{-7}$ kg/m$^3$ for 1s integration time. This $10^{-10}$ relative sensitivity represents 3 orders of magnitude improvement over established techniques such as magnetic densimeters[34]. Finally, Fig. 5c focuses on the most viscous liquid. At high viscosity, the assumption that viscous and acoustic losses are independent breaks down and experimental results are better explained by the viscous model alone. The validity of analytical formulas in the most viscous liquids allows again to evaluate the sensitivity to viscosity changes, leading a relative sensitivity of $(\delta\mu/\mu)_{min}=5\times10^{-12}$ for the same disk as above. These ultimate mass and rheological sensitivities, quoted at the thermodynamical limit, clearly show the potential of scaling down microfabricated optomechanical resonators to the nanoscale. The smallest disks experimented here reach a sub-μm$^3$ mechanical and optical volume, but further miniaturization may lead to even more stunning performances.

In this work, optomechanical techniques allow resolving the Brownian motion of nanoscale resonators in liquids, paving the way to low-noise optical driving of GHz motion through radiation pressure and electrostriction. We have used the precision of nano-optomechanical techniques to validate original hydrodynamical models at very high frequency. While putting miniature disk fluidic sensors on a firm ground, this also provides a solid picture of nano-optomechanical dissipation in liquids. This will be of importance for applications to come, ranging from quantum-enhanced technologies in aqueous environments[35] to high-speed optomechanical imaging probes. The concept of nano-optomechanical disks in liquids hence opens several routes, which will all benefit from the mature technological control of semiconductors, offering optoelectronic integration, lab-on-chip protocols and massively parallel architectures.

**Methods**

*Experiments:* Light from an external tunable cavity diode laser fiber-guided, polarization-controlled, injected into the chip's waveguide using a micro-lensed fiber, collected at the guide's output with a microscope objective and sent on a photodiode. The laser wavelength is first swept to acquire the optical spectrum of the disk and then set onto the blue flank of a WGM resonance. In this case the mechanical information is imprinted onto the radio-frequency spectrum of light and revealed by a spectrum analyzer plugged at the photodiode output.

*Nanofabrication:* The disks are fabricated out of a GaAs (320 nm)/$Al_{0.8}Ga_{0.2}As$ (2.000 nm)/GaAs wafer grown by molecular beam epitaxy. Disks are positioned in the vicinity of GaAs tapered suspended waveguides to allow evanescent optical coupling. Disks and waveguides are patterned in a resist mask by electron beam lithography and dry-etched by Inductively Coupled Plasma Reactive Ion Etching (ICP-RIE) with a mixture of $SiCl_4$ and Ar plasmas. The AlGaAs sacrificial layer is selectively under-etched by hydrofluoric acid (HF) to form the pedestals selective. A diluted HF:$H_2O$ solution (1.22 % in volume) at 4°C is combined with a slow agitation in the solution to reach pedestal radius below 200 nm.

*Liquids:* The 3 liquids employed in experiments are inert non-ionic perfluorinated solutions from Galden (HT 170, HT 230 and HT 270).

 **Acknowledgments**

E. Gil-Santos and D. T. Nguyen acknowledge support by the Research in Paris program of the Ville de Paris and by the ANR through the NOMADE Project respectively. E. Gil-Santos, C. Baker, W. Hease and I. Favero acknowledge support by the ERC through the GANOMS project.

**Author's contributions**

E. G and I. F conceived and designed the experiments, and developed the models. C.B, D.T. N and W. H contributed in fabrication and experimental techniques. A. L. grew the epitaxial material. E. G. performed systematic experiments. All authors discussed the results and wrote the paper.